\documentclass[10pt,conference]{IEEEtran}
\IEEEoverridecommandlockouts

\usepackage{cite}
\usepackage{amsmath,amssymb,amsfonts}
\usepackage{algorithmic}
\usepackage{graphicx}
\usepackage{textcomp}
\usepackage{xcolor}
\usepackage{xspace}
\usepackage{booktabs}
\usepackage{hyperref}
\usepackage{tcolorbox}
\usepackage{lipsum} \usepackage{enumitem}
\usepackage{multirow}
\usepackage{siunitx}
\usepackage{array}
\usepackage[all]{nowidow}

\def\BibTeX{{\rm B\kern-.05em{\sc i\kern-.025em b}\kern-.08em
    T\kern-.1667em\lower.7ex\hbox{E}\kern-.125emX}}

\newcommand{\lintq}{LintQ\xspace}
\newcommand{\approach}{LintQ-LLM\xspace}
\newcommand{\approachCoT}{LintQ-LLM+CoT\xspace}
\newcommand{\approachRAG}{LintQ-LLM+RAG\xspace}

\newcommand{\etal}{et al.\xspace}

\begin{document}

\title{Beyond Rules: LLM‑Powered Linting for Quantum Programs}

\makeatletter
\newcommand{\linebreakand}{\end{@IEEEauthorhalign}
  \hfill\mbox{}\par
  \mbox{}\hfill\begin{@IEEEauthorhalign}
}
\makeatother

\author{
\IEEEauthorblockN{Pietro Cassieri}
\IEEEauthorblockA{\textit{Department of Computer Science} \\ \textit{University of Salerno} \\
Salerno, Italy \\
pcassieri@unisa.it}
\and
\IEEEauthorblockN{Giuseppe Scanniello}
\IEEEauthorblockA{\textit{Department of Computer Science} \\ \textit{University of Salerno} \\
Salerno, Italy \\
gscanniello@unisa.it}
\and
\IEEEauthorblockN{Seung Yeob Shin}
\IEEEauthorblockA{\textit{University of Luxembourg} \\
Luxembourg, Luxembourg \\
seungyeob.shin@uni.lu}
\and
\linebreakand
\IEEEauthorblockN{Fabrizio Pastore}
\IEEEauthorblockA{\textit{University of Luxembourg} \\
Luxembourg, Luxembourg \\
fabrizio.pastore@uni.lu}
\and
\IEEEauthorblockN{Domenico Bianculli}
\IEEEauthorblockA{\textit{University of Luxembourg} \\
Luxembourg, Luxembourg \\
domenico.bianculli@uni.lu}

}

\maketitle

\begin{abstract}

As quantum computing transitions from theoretical experimentation to its practical application, the reliability of quantum software has become a critical bottleneck. Traditional static analysis techniques for quantum programs---primarily rule-based linters---are increasingly inadequate; they struggle to keep pace with rapidly evolving APIs and fail to capture complex, context-dependent quantum programming problems. This results in high maintenance overhead and limited detection capabilities. In this paper, we introduce \approachCoT and \approachRAG, novel approaches that redefine the detection of quantum programming problems by employing Large Language Models (LLMs) specialized, respectively, via Chain-of-Thought (CoT) prompting and a Retrieval-Augmented Generation (RAG) system that grounds the model's reasoning in a curated knowledge base of verified quantum programming problems and best practices. 
We conducted a rigorous and manual comparative evaluation against the state-of-the-art rule-based tool, LintQ, using a corpus of 55 Qiskit programs. Our results show that LLM-based approaches, with and without RAG, outperform LintQ in terms of quantum programming problems detection correctness (precision) and completeness (recall).
Overall, LLM-based approaches were more effective than \lintq~(F1-score equal to 0.70 and 0.68 vs. 0.41). 
Furthermore, the RAG-enhanced variant demonstrated a slightly superior precision, effectively reducing false positives. Our findings suggest that LLMs provide a scalable and adaptive foundation for the next generation of linters in quantum software engineering.

\end{abstract}

\section{Introduction}

Ensuring the reliability of quantum software has become a paramount challenge as quantum computing transitions from a theoretical pursuit to a practical engineering discipline. In this evolving landscape, developers increasingly rely on tools like Qiskit~\cite{Qiskit} to manipulate both quantum and classical bits. However, the unique principles of quantum mechanics---such as superposition, entanglement, and the collapse of quantum states upon measurement---introduce specialized programming ``anti-patterns'' or bugs that traditional classical software analysis tools are often ill-equipped to detect.

Recent advances in the quantum research field have identified ten common, quantum-specific, programming problems, categorized into families such as gate-related issues, resource allocation errors, and implicit API constraint violations. To address these, state-of-the-art static analysis frameworks like \lintq~\cite{PaltenghiP24} have been developed. While \lintq~achieves notable precision by utilizing declarative queries over formal abstractions of quantum concepts, it relies heavily on manually crafted deterministic rules~\cite{PaltenghiP24}. This reliance creates a significant maintenance burden and limits the tool's ability to adapt to the rapidly evolving nature of quantum APIs~\cite{shin2025}.

In previous work~\cite{shin2025}, some of the authors of this paper introduced the foundational concepts for \approach, an approach that leverages Large Language Models (LLMs) to automate the detection of quantum programming problems in Qiskit programs. This approach was limited to single-prompt interactions and did not fully utilize the reasoning capabilities of the models. 

In this paper, we improve \approach by introducing a multi-prompt Chain-of-Thought (CoT) interaction designed to guide the model through strategic planning, code understanding, and localizing the problems. By moving away from rigid rules and toward the flexible reasoning capabilities of LLMs, this enhanced approach, named \approachCoT, provides a more adaptable solution to detect programming problems. 
In addition, we also propose \approachRAG, a variant of \approachCoT that uses a Retrieval-Augmented Generation (RAG) extension to further bridge the gap between flexible reasoning and formal correctness. The overarching goal of this variant is to ground the model's reasoning in a specialized knowledge base of manually verified true positive instances (i.e., actual quantum programming problems). By providing the model with ``one-shot'' learning examples, with semantic similarity to the analyzed code, we aim to improve the correctness of the analysis and reduce the hallucination of false positives (FPs).

We also conducted a comprehensive empirical evaluation of both \approachCoT and \approachRAG in terms of detection correctness (precision), completeness (recall), and effectiveness (F1-score) of quantum programming problems.
Specifically, we performed a manual, rigorous comparative evaluation of \approachCoT and \approachRAG 
against the state-of-the-art rule-based tool, \lintq, using a corpus of 55 Qiskit programs, including both real-world code and synthetic fault-injected files. 

Our results show that \approachCoT, with and without RAG, outperforms \lintq~in terms of correctness and completeness.
Also, the effectiveness of the LLM-based approaches is higher than that of \lintq~(0.70 and 0.68 vs. 0.41). Furthermore, the RAG-enhanced variant yielded a slightly superior precision, effectively reducing FPs. 
In summary, our results suggest that LLMs provide a scalable and adaptive foundation for the next generation of quantum software quality assurance tools (e.g., linters).

To summarize, this paper makes the following contributions:
\begin{itemize}
    \item A multi-prompt CoT pipeline for \approach, namely \approachCoT, which enhances the foundational LLM-based linting concept introduced in prior work~\cite{shin2025}, as well as \approachRAG, a variant of \approachCoT that uses RAG to ground reasoning in verified quantum programming problem examples.
    \item  A rigorous empirical assessment of \lintq, \approachCoT, and \approachRAG in terms of effectiveness as well as correctness and completeness of the detected quantum programming problems. 
\end{itemize}

\section{Background and Related Work} \label{sec:RelatedWork}

This section establishes the foundational concepts and related literature framing our work.

\subsection{Quantum Software Testing and Analysis}

Several automated approaches have been proposed to ensure the correctness of quantum compilers and simulators. One prominent example is QDiff~\cite{wangQDiffDifferentialTesting2021}, which adapts differential testing to quantum software. It generates semantically equivalent program variants through source-to-source transformations and compares their outputs across different back-ends and optimization levels using statistical tests such as the Kolmogorov–Smirnov test. Similarly, MorphQ~\cite{paltenghiMorphQMetamorphicTesting2023} applies metamorphic testing to the Qiskit toolchain by transforming circuits---for instance, by changing basis gates or converting to and from OpenQASM---and checking whether expected relationships between outputs hold. More recently, Fuzz4All~\cite{xiaFuzz4AllUniversalFuzzing2024} explores universal fuzzing for quantum systems by leveraging LLMs to generate and mutate diverse, syntactically valid programs that can stress-test quantum platforms.

Testing quantum software poses challenges distinct from those in classical computing, as highlighted by Miranskyy et al.~\cite{miranskyyYourQuantumProgram2020}. Quantum programs are inherently probabilistic, quantum states cannot be copied due to the no-cloning theorem, and any measurement irreversibly collapses the state. Together, these properties make it difficult to observe, reproduce, and validate program behavior, complicating the entire testing process. Testing has largely focused on dynamic techniques that can handle probabilistic outputs. Quito, proposed by Ali et al.~\cite{aliAssessingEffectivenessInput2021a}, introduces input and output coverage criteria combined with statistical oracles, such as the Wilcoxon signed-rank test, to assess whether a test passes or fails over multiple executions. In a similar vein, QuanFuzz~\cite{wangPosterFuzzTesting2021} uses search-based techniques, including genetic algorithms, to generate inputs that maximize coverage of quantum-relevant behaviors, particularly around measurement operations.

Researchers have also proposed runtime assertion mechanisms to address the intrinsic difficulty of inspecting intermediate quantum states. For example, Huang and Martonosi~\cite{huangStatisticalAssertionsValidating2019} introduced statistical assertions based on Chi-square tests to check whether observed measurement distributions match expected outcomes. Differently, Li et al.~\cite{liProjectionbasedRuntimeAssertions2020} proposed Proq, a projection-based approach grounded in Birkhoff–von Neumann quantum logic. By representing predicates as projections onto subspaces, Proq enables partial verification of intermediate states using local measurements and ancilla qubits, reducing the risk of collapsing the entire quantum state.

Despite their shown effectiveness, dynamic approaches generally require repeated program execution. This can be costly in terms of time and resources, especially when running on noisy intermediate-scale quantum (NISQ) hardware, where results are affected by noise and limited qubit fidelity. These limitations have motivated increasing interest in static analysis techniques.
In fact, static analysis avoids execution altogether, sidestepping both measurement-related issues and hardware constraints. For example, QSmell~\cite{ChenCCSA23} identifies quantum-specific code smells—such as overly long circuits or misaligned qubit mappings—that may degrade performance or reliability. QChecker~\cite{ZhaoWLZ23} focuses on problem detection by constructing specialized abstract syntax trees (ASTs) and matching extracted properties against known problem patterns from the Bugs4Q benchmark. Meanwhile, QCPG~\cite{KaulKB23} extends classical code property graphs to the quantum domain, integrating ASTs, control-flow graphs, and program-dependence graphs to model quantum constructs like qubits, gates, and measurements, enabling problem detection through graph queries. Unlike these dynamic and rule-based static approaches, our work leverages LLM to perform quantum linting. By employing multi-prompt CoT reasoning and RAG, we provide a more flexible approach that generalizes across evolving APIs without requiring manually crafted rules.

\begin{table}[t]
\caption{Quantum-specific programming problems and their descriptions.}\label{tab:lintq_problems}
\centering
\begin{tabular}{l p{0.65\columnwidth}}
\toprule
\textbf{Problem} & \textbf{Description} \\
\midrule
\multicolumn{2}{l}{\textit{Measurement- or Gate-related Problems}} \\
DoubleMeas & Two consecutive measurements are performed on the same qubit state. \\
OpAfterMeas & A gate is applied to a qubit after it has already been measured. \\
MeasAllAbuse & Measurement results are stored in a newly and implicitly created register, despite the presence of an existing classical register. \\
CondWoMeas & A conditional gate is applied without measuring the associated register. \\
ConstClasBit & A qubit is measured without undergoing any prior transformation. \\
\midrule
\multicolumn{2}{l}{\textit{Resource Allocation Problems}} \\
InsuffClasReg & There are not enough classical bits to store the measurement results of all qubits. \\
OversizedCircuit & The quantum register includes qubits that remain unused. \\
GhostCompose & Two circuits are composed, but the resulting composed circuit is not utilized. \\
\midrule
\multicolumn{2}{l}{\textit{Implicit API Constraint Violations}} \\
OpAfterOpt & A gate is applied to the circuit after transpilation. \\
OldIdenGate & An identity gate is created using an API that has been removed. \\
\bottomrule
\end{tabular}
\end{table}

\subsection{Quantum Problem Detection: LintQ}

Quantum programming problems frequently arise when developers misuse quantum constructs or misunderstand the underlying quantum properties. Recently, studies have cataloged these issues to aid in the development of automated analysis tools. In particular, ten common quantum-specific programming problems have been identified and categorized into three main families: measurement- or gate-related problems, resource allocation problems, and implicit API constraint violations~\cite{PaltenghiP24}. In Figure~\ref{fig:quantum_problem}, we show an example of the OpAfterMeas problem, while Table~\ref{tab:lintq_problems} summarizes these quantum programming problems (presented in the \lintq work~\cite{PaltenghiP24}) by providing a short description for each of them.

\begin{figure}[]
    \centering
    \includegraphics[width=0.7\columnwidth]{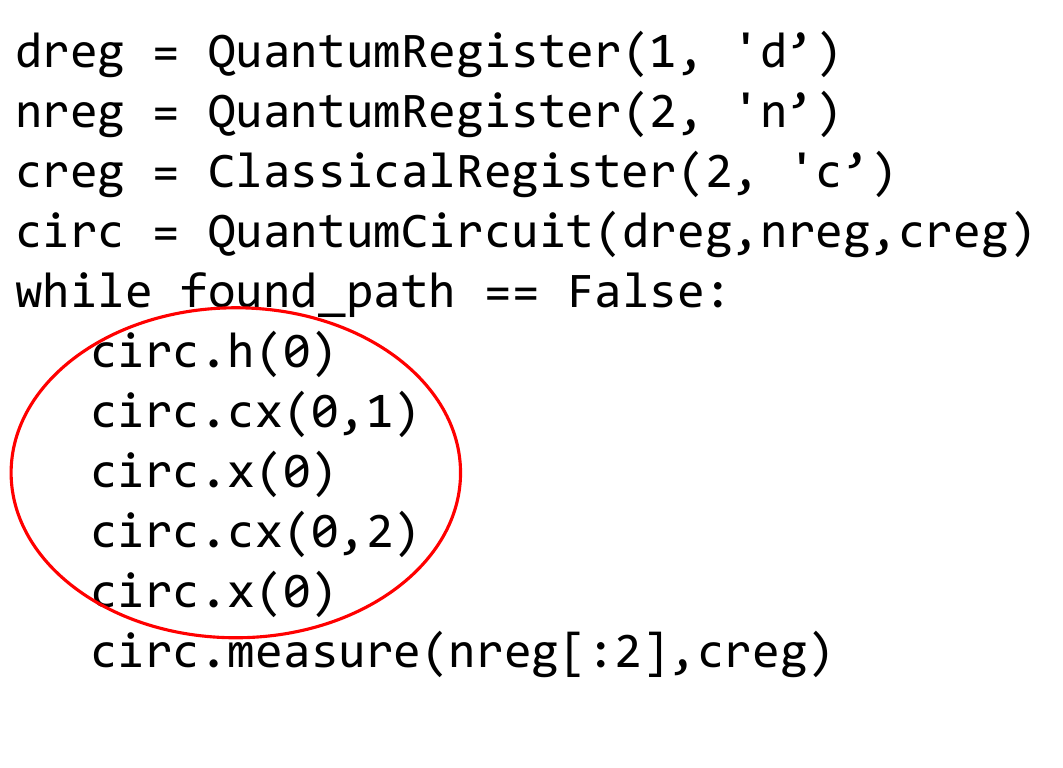}
    \caption{An example of code affected by \textit{OpAfterMeas}. The portion of the quantum programming affected by this problem is highlighted in the red ellipse: starting from the second iteration of the while loop, quantum gates are applied to qubits that have already been measured.}
    \label{fig:quantum_problem}
\end{figure}

LintQ~\cite{PaltenghiP24} is the state-of-the-art static analysis tool specifically designed to detect quantum-specific problems in source code written using Qiskit~\cite{Qiskit}. It introduces a formal set of abstractions representing common quantum computing concepts, such as quantum registers, classical registers, quantum circuits, gates, qubit usage, and measurements. These abstractions provide a high-level representation of the program, enabling LintQ to perform robust static analyses rapidly without needing to thoroughly process the lower-level Python implementation details.

Built upon these abstractions, LintQ leverages CodeQL \cite{avgustinovQLObjectorientedQueries2016}, a general-purpose, robust static analysis engine, to execute declarative queries over the behavioral representation of the Qiskit code. Each analysis corresponds to a query verifying against the ten problems described in Table~\ref{tab:lintq_problems}. A previous empirical evaluation of LintQ on 7568 real-world Qiskit programs achieved an overall precision of 62.5\%~\cite{PaltenghiP24}. While this demonstrated the feasibility and effectiveness of query-based problem detection, LintQ inherently relies on manually crafted deterministic patterns and rules. This architectural choice demands significant manual effort to maintain and limits the tool's adaptability to evolving quantum programming practices, thus highlighting the need for novel, more adaptable solutions to quality assurance in quantum software.

\subsection{LLM-based Quantum Problem Detection: \approach}
\label{subsec:Approach}
To overcome the rigid nature of rule-based linters, prior work introduced \approach~\cite{shin2025}, an LLM-based linting approach designed to identify quantum programming problems in Qiskit source code. The core logic to analyze a Qiskit code fragment relies on querying an LLM independently for each type of problem, ensuring high specificity while strictly respecting the operational context window of the model. Figure~\ref{fig:lintq-LLM_overview} illustrates the overall architecture of \approach. \approach analyzes one file at a time and prepares a separate prompt for each type of problem: in each prompt, it inserts only the instructions related to that check (i.e., a description of the problem and an example) along with the code to be analyzed. This keeps the context shorter, the code occupies more space in the token budget, and the model does not have to handle so many instructions at once, reducing the risk of confusion between the various constraints.

\begin{figure}[t]
    \centering
    \includegraphics[width=\columnwidth]{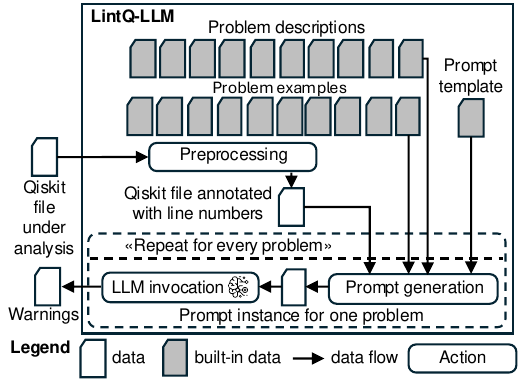}
    \caption{Architecture and data flow overview of \approach.}
    \label{fig:lintq-LLM_overview}
\end{figure}
The problem detection strategy in \approach relies on a single-prompt interaction. A zero-shot prompt is constructed to instruct the model to act as a source code linter with expert knowledge in quantum software. The prompt specifies the exact problem to detect, including its name, a detailed description, a generic example of the problem, and embeds the target source code to be analyzed. Finally, it mandates that the response must be a single JSON object.
While \approach demonstrated the foundational feasibility of employing LLMs for quantum code analysis, the single-prompt mechanism structurally can limit the reasoning depth and context awareness of the model, occasionally resulting in FPs when analyzing complex quantum circuits. This represents the main motivation behind the definition of \approachCoT and \approachRAG, both introduced in the next section. 

\section{\approachCoT and \approachRAG} \label{sec:Approach}

To enhance the performance of \approach, we propose \approachCoT and \approachRAG, which significantly evolve the LLM interaction mechanism through CoT and incorporate RAG.

\subsubsection{Multi-prompt Chain-of-Thought Pipeline}
\label{subsubsec:multiprompt}

The problem detection strategy is implemented through a multi-prompt CoT interaction, which consists of one system prompt and two sequential user prompts (note that despite being called user prompts, their generation is automated by our approach). To ensure optimal structural layout and instruction clarity, these prompts were systematically generated and refined during development, leveraging the OpenAI Prompt Optimizer~\cite{openai_prompt_optimizer}. 
The overall structure of our prompting mechanism is shown in Figure~\ref{fig:prompt} and summarized below.

The system prompt instructs the model to act as a source code linter with expert knowledge in quantum software and mandates that its response must be a single JSON object.

The first automated user prompt provides a comprehensive task definition: it specifies the exact problem to detect, including its name, a detailed description, and a concrete code example of the quantum programming problem (dynamically provided by the RAG module in \approachRAG). Specifically, this prompt guides the LLM through a multi-step reasoning process:
\begin{enumerate}
    \item \textbf{Strategic Planning:} Formulate a ``Detection Strategy'' outlining the conceptual plan to identify the problem, including primary API elements and logical checks.
    \item \textbf{Code Understanding:} Create a ``Code Summary'' to briefly describe the essential components and operations within the source code.
    \item \textbf{Problem Detection Logic:} Apply the strategy to inspect the code and identify all instances of the problem.
    \item \textbf{Report Results:} For each detected case, generate a JSON object containing the exact code ``snippet'', an array of ``lines'' numbers, and a detailed explanation. If no problems are found, an empty JSON object is returned.
\end{enumerate}

The second automated user prompt delivers the line-numbered source code, instructing the model to perform the analysis based on the strategy established in the previous turn.

\begin{figure*}[htbp]
    \centering
    \includegraphics[width=\textwidth]{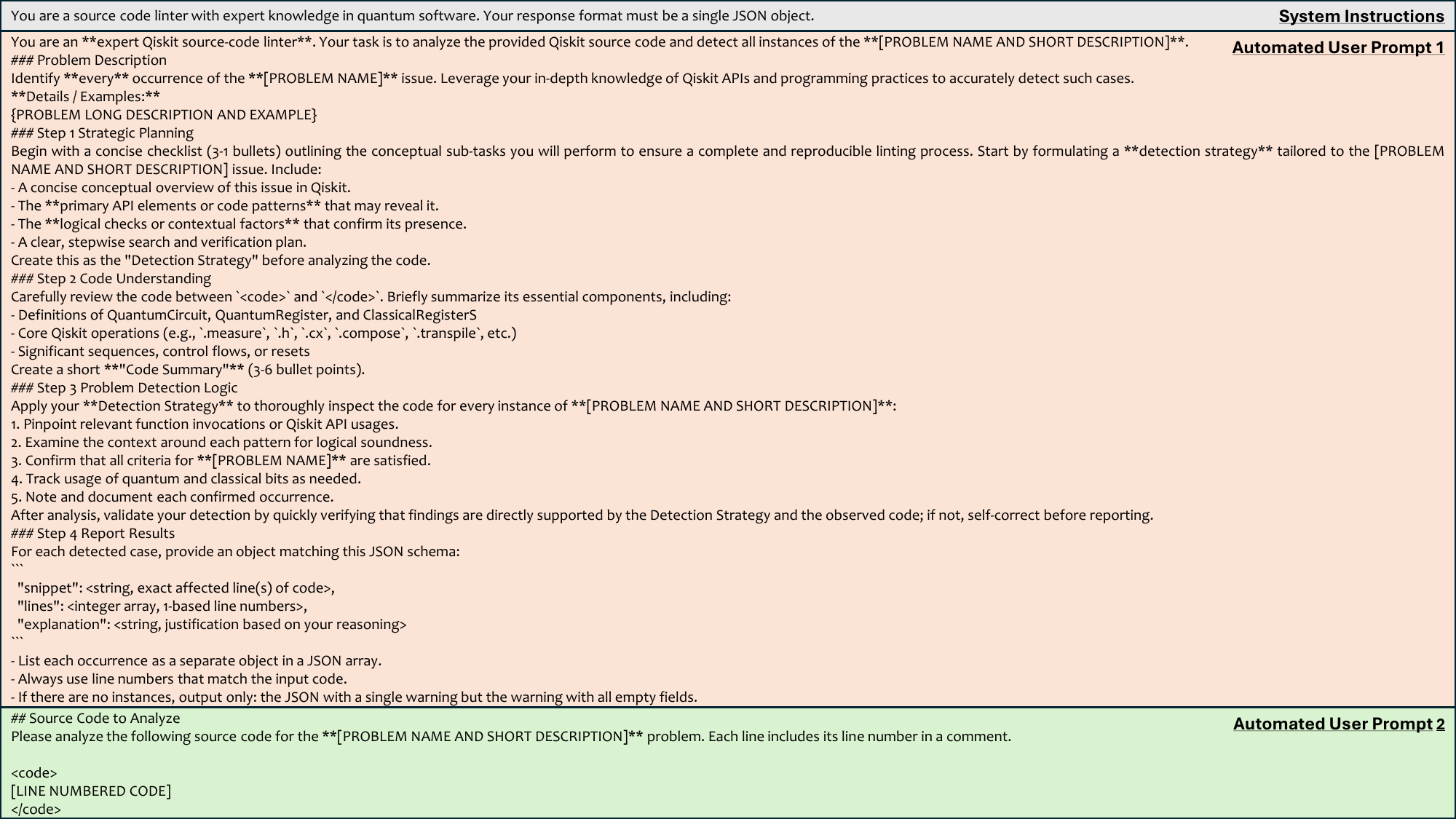}  
    \caption{The multi-prompt structure used by \approachCoT and \approachRAG. The interaction consists of a system prompt and two sequential automated user prompts designed to first establish a detection strategy with examples and then analyze the target source code.}
    \label{fig:prompt}
\end{figure*}

\subsubsection{Knowledge base creation}
\approachRAG incorporates a RAG system, leveraging a specialized vector database as its knowledge retrieval backend. The creation of this knowledge base followed a rigorous inclusion and exclusion protocol operating on the dataset previously evaluated by the original \lintq authors~\cite{PaltenghiP24}. Figure~\ref{fig:rag_generation} depicts the process responsible for building and indexing this knowledge base. 

\begin{figure}[htbp]
    \centering
    \includegraphics[width=0.9\columnwidth]{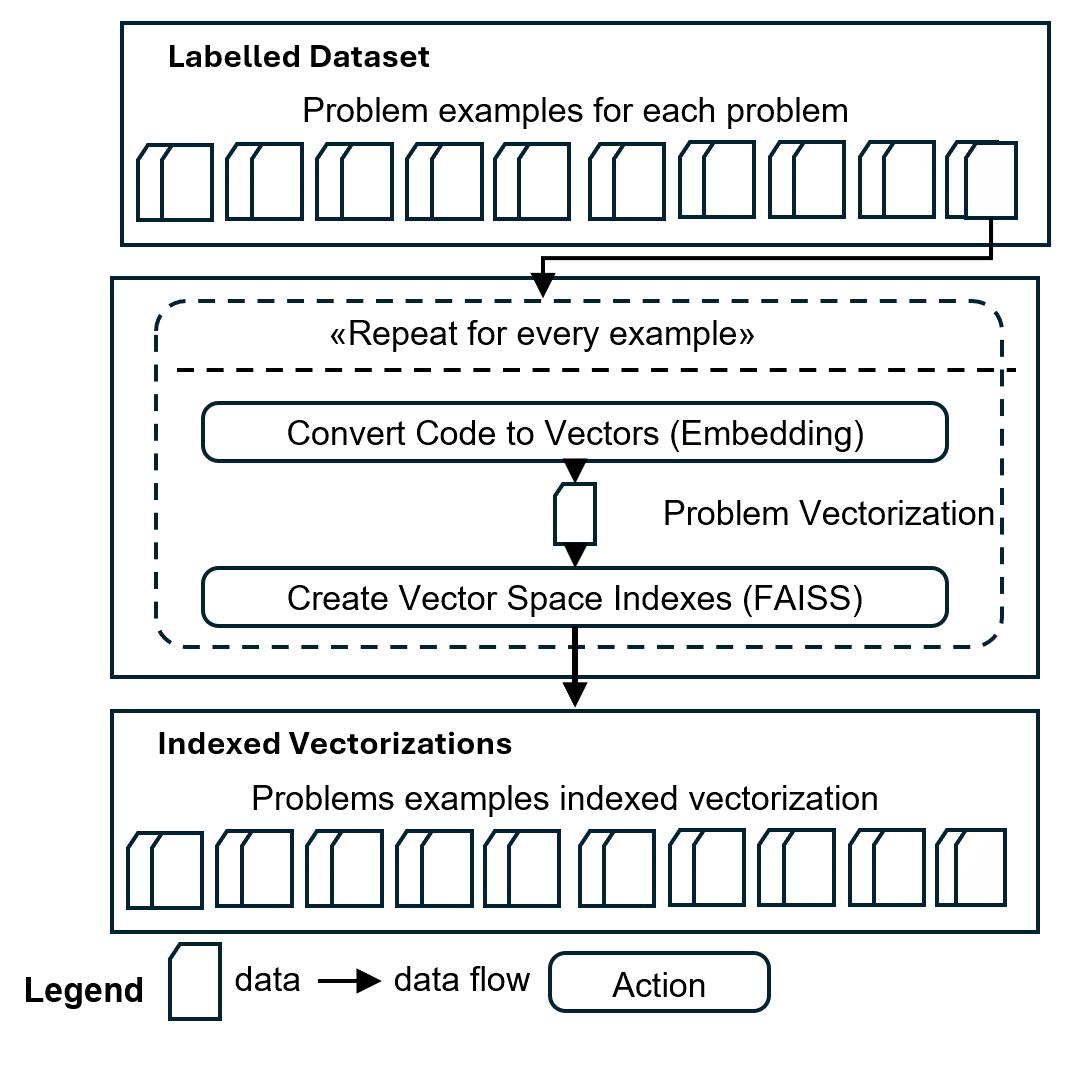}
    \caption{Knowledge Base creation process for the RAG integration.}
    \label{fig:rag_generation}
\end{figure}

The foundation for the valid context injections is composed of manually verified True Positives (TPs). From a set of 7568 Qiskit files, Paltenghi~\etal obtained 4699 warnings of quantum-related problem detection~\cite{PaltenghiP24}. Of this 4699 warnings, they manually analyzed 345 warnings and annotated the warnings as TPs (True Positives), FPs (False Positives), or NW (Noteworthy, used for borderline cases). Although the meaning of TP and FP is well recognized, please note that they are formally defined in Section~\ref{subsec:metrics}. At the end of the manual labeling, they obtained 165 TP instances~\cite{PaltenghiP24}. 
We excluded FPs and NW cases from our knowledge base
to ensure that the retrieval system grounds its reasoning strictly on unambiguous examples. 
We programmatically annotated the 165 files with the TP selected for our knowledge base:
the lines explicitly causing the warning were commented with a standardized label \texttt{\# Problem: <Problem Description>}.

After preparing the source files, we systematically embedded all candidate examples using OpenAI's \texttt{text-embedding-3-large} model~\cite{text-embedding-3}. This model ensures optimal semantic compatibility with the GPT-5 model family ultimately responsible for the reasoning step. To avoid exceeding the strict context window limit during execution, we filtered the selected files using the \texttt{tiktoken} library~\cite{tiktoken}: examples with more than 8192 tokens were excluded, resulting in the removal of eight files. The remaining subset corresponds to a knowledge base of 157 files. These files were sequentially vectorized and stored in a Facebook AI Similarity Search (FAISS) index~\cite{douze2025faisslibrary} capable of determining similarity through Euclidean distance.
The ultimate composition of the RAG database varies in available examples per problem type, distributed as reported in Table~\ref{tab:rag_composition}.

\begin{table}[t]
    \centering
    \caption{Composition of the RAG Knowledge Base by number of examples per Quantum problem Type.}
    \label{tab:rag_composition}
    \begin{tabular}{lc|lc}
        \toprule
        \textbf{Quantum problem} & \textbf{Count} & \textbf{Quantum problem} & \textbf{Count} \\
        \midrule
        OpAfterMeas & 38 & DoubleMeas & 15 \\
        ConstClasBit & 25 & CondWoMeas & 14 \\
        OversizedCircuit & 20 & OldIdenGate & 10 \\
        InsuffClasReg & 18 & GhostCompose & 6 \\
        MeasAllAbuse & 16 & OpAfterTransp & 4 \\
        \bottomrule
    \end{tabular}
\end{table}
\subsubsection{Retrieval mechanism}
In the \approachRAG execution phase, dynamically analyzing a target file translates into a comprehensive nearest-neighbor search paired with one-shot learning. As depicted in Figure~\ref{fig:rag_implementation}, this approach evolves the \approach architecture from an ungrounded inference model to a dynamically parameterized static analyzer. The integration of RAG into \approachRAG augments each LLM query with a retrieved example selected from the indexed knowledge base. Given the input Qiskit file, a preprocessing step annotates the code (e.g., with line numbers) and prepares it for analysis, while in parallel the system retrieves the most relevant problem example based on similarity with the current input.

\begin{figure}[htbp]
    \centering
    \includegraphics[width=0.9\columnwidth]{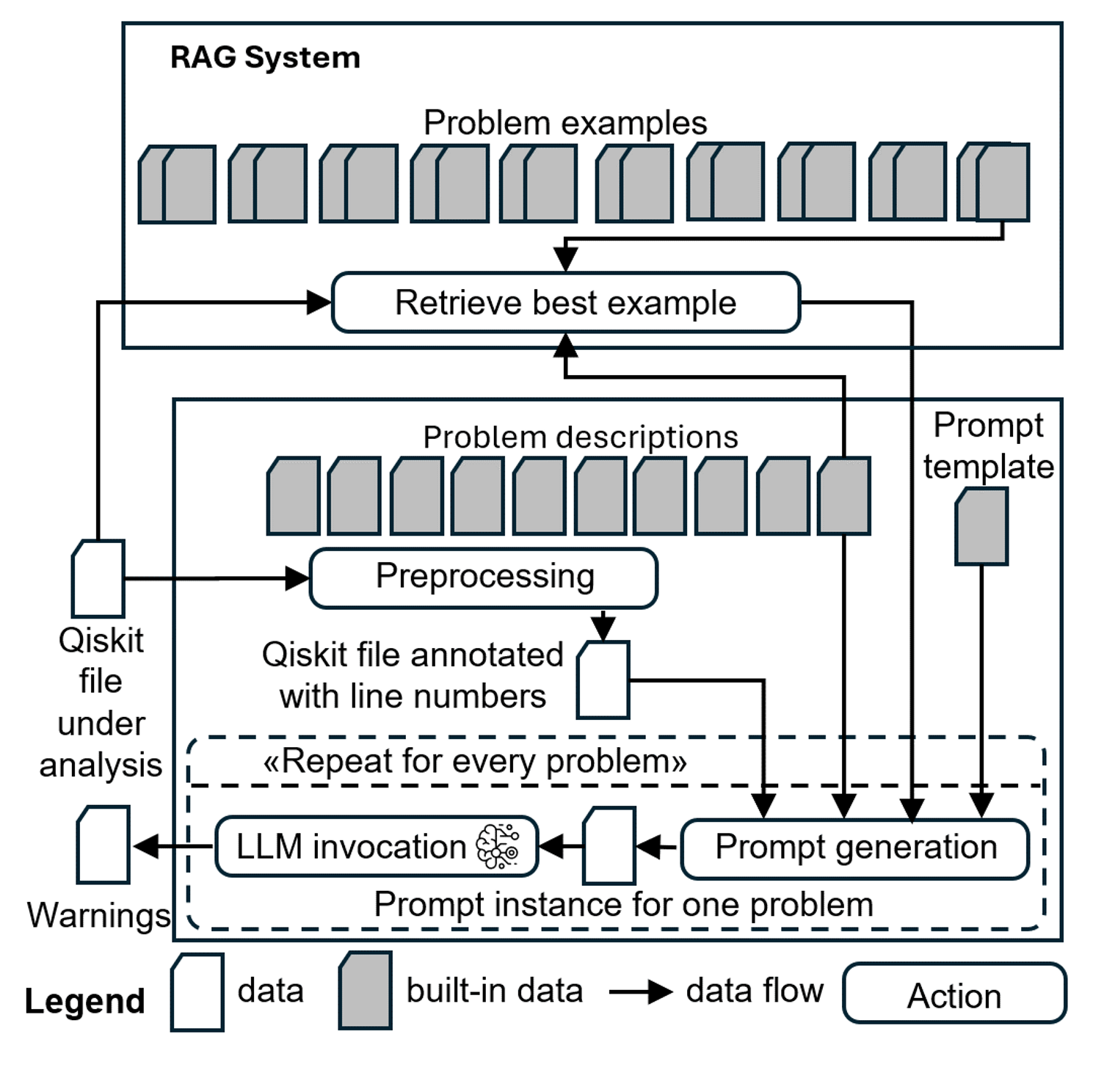}
    \caption{Retrieval mechanism integration mapped in \approachCoT.}
    \label{fig:rag_implementation}
\end{figure}

When processing an incoming Qiskit program, its source code is initially tokenized to verify its bounds against the embedding constraint window. Assuming the valid 8192 bounds stand, the \texttt{text-embedding-3-large} model builds a vector representation of the current file. Due to the domain specificity of distinct problem types, the system operates across separate FAISS indices designed independently for each of the ten investigated problems.

For a specific quantum programming problem analysis, the framework queries the aligned FAISS index, recovering the single most semantically similar verified file ($k=1$). By retrieving a known, mathematically comparable instance containing the explicit problem, the \approachRAG architecture dynamically structures a reference snippet. This snippet, coupled with its embedded description of the problem's exact location, is then injected into the prompt. Consequently, this guides the final LLM validation by grounding its judgment in a high-quality, context-aware reference example retrieved from the FAISS knowledge base.

\section{Empirical Study Design}
\label{sec:study_design}

Our overarching goal is to assess whether shifting from rule-based static analysis to LLM-driven reasoning provides benefits in the detection of quantum problems in Qiskit programs. Given this goal, we formulate the following two research questions (RQs):
\begin{itemize}

\item \textbf{RQ1:} \textit{How do \approachCoT and \approachRAG compare to \lintq in terms of effectiveness, correctness, and completeness in detecting quantum programming problems?}\\
It focuses on a comparative evaluation against the current state of the art, aiming to determine whether the proposed LLM-based techniques represent a meaningful advancement over traditional rule-based static analysis.

\item \textbf{RQ2:} 
\textit{To what extent does \approachRAG enhance \approachCoT's ability to detect quantum programming problems with respect to correctness, completeness, and effectiveness?}\\
This RQ narrows the focus to the individual LLM-based approaches, specifically investigating the contribution of RAG (if any) in improving detection performance (correctness, completeness, and effectiveness). 
\end{itemize}

The rest of this section describes the experimental setup, the construction of the evaluation corpus, and the protocol followed to assess the efficacy of \approachCoT and \approachRAG. Figure~\ref{fig:study_design_flow} provides an overview of the used design.

\begin{figure*}[htbp]
    \centering
    \includegraphics[width=1\textwidth]{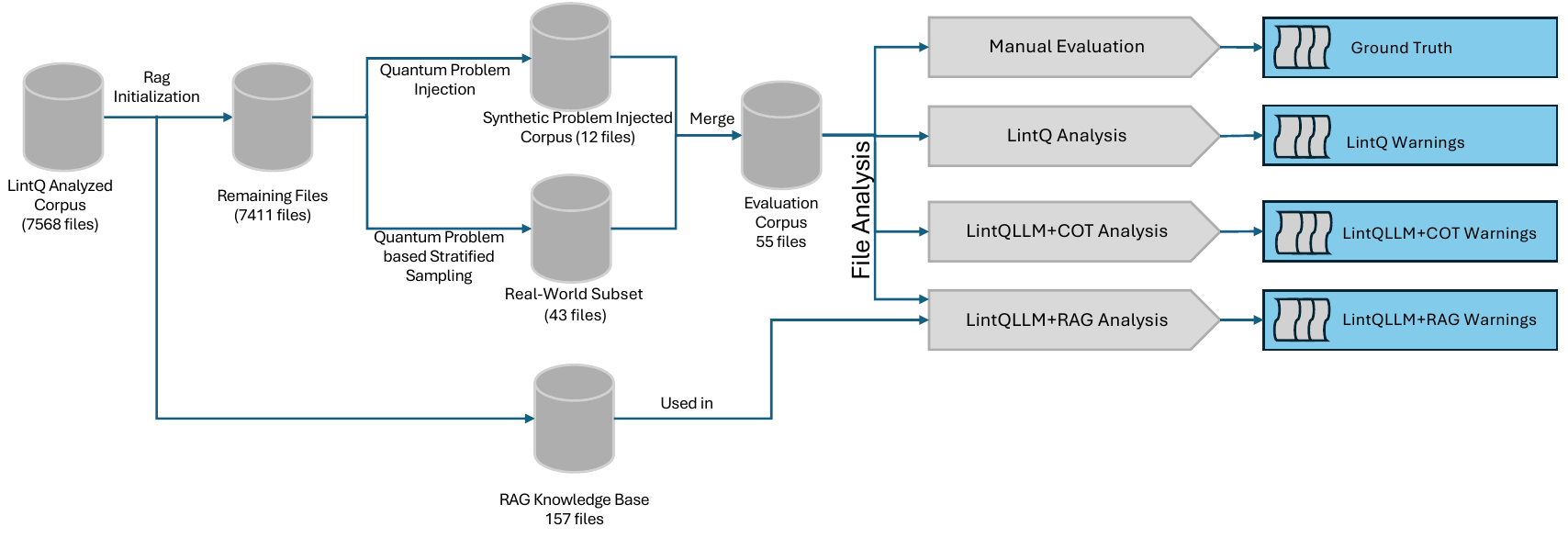}
    \caption{Experimental design for the construction and validation of the Evaluation Corpus.}
    \label{fig:study_design_flow}
\end{figure*}

\subsection{Evaluation Corpus Construction}
\label{subsec:corpus_construction}

To address our RQs, we constructed a balanced stratified evaluation corpus. The process began with the original LintQ corpus consisting of 7568 Qiskit files. We first excluded the 157 files manually verified by Paltenghi~\etal \cite{PaltenghiP24} and already incorporated into the RAG knowledge base. The remaining subset, consisting of 7411 files, was further filtered to include only those with a token count within the 8192 limit of the chosen embedding model, ensuring that the RAG retrieval mechanism could operate on the entire file content.

From this filtered subset, we performed a stratified selection focused on {potential quantum problems} identified by unverified LintQ reports. Our objective was to curate a balanced dataset of {55 files}: up to five files for each of the ten supported quantum problem categories and five files flagged as clean (zero warnings) to serve as a control group. At this stage, the actual presence of the problems was unknown, as the selection relied on the original tool's unverified output.

For categories where the filtered subset contained fewer than five eligible files (i.e., extremely rare problem types), we utilized problem injection~\cite{NatellaInjection}. We manually introduced specific quantum problems into verified clean files following the patterns documented by Paltenghi~\etal \cite{PaltenghiP24}. Concretely, we first analyzed the formal definitions and example instances of each problem provided in the \lintq work~\cite{PaltenghiP24}, and then reproduced those patterns by modifying existing circuits that did not originally exhibit the problem. This process resulted in a final Evaluation Corpus composed of 43 real-world files and 12 synthetic (injected) files, as detailed in Table~\ref{tab:evaluation_corpus}. Some of the files selected had multiple instances of different LintQ warnings; in fact, at the end of the sampling, and after the analysis of the synthetic files with LintQ, we had a total of 77 warnings across the 55 files.

\begin{table}[t]
  \centering
  \caption{Composition of the Evaluation Corpus across all quantum problem categories.}
  \label{tab:evaluation_corpus}
  \small
  \setlength{\tabcolsep}{8pt}
  \begin{tabular}{lccc}
    \toprule
    \textbf{Quantum Problem} & \textbf{Real Files} & \textbf{Injected Files} & \textbf{Total} \\
    \midrule
    Clean Files        & 5 & 0 & 5 \\
    CondWoMeas         & 3 & 2 & 5 \\
    ConstClasBit       & 5 & 0 & 5 \\
    DoubleMeas         & 5 & 0 & 5 \\
    GhostCompose       & 1 & 4 & 5 \\
    InsuffClasReg      & 5 & 0 & 5 \\
    MeasAllAbuse       & 4 & 1 & 5 \\
    OldIdenGate        & 5 & 0 & 5 \\
    OpAfterMeas        & 5 & 0 & 5 \\
    OversizedCircuit   & 5 & 0 & 5 \\
    OpAfterTransp      & 0 & 5 & 5 \\
    \midrule
    \textbf{Total}     & \textbf{43} & \textbf{12} & \textbf{55} \\
    \bottomrule
  \end{tabular}
\end{table}

\subsection{Experimental Procedure}
\label{subsec:procedure}
The effectiveness of the proposed approaches was evaluated through four parallel identification actions performed on the Evaluation Corpus:
\begin{enumerate}
    \item \textbf{Manual Validation}: A human annotator (the first author) conducted a line-by-line analysis of all 55 files to identify every occurrence of each quantum problem, establishing the {Ground Truth} for the study.
    \item \textbf{Baseline Tool Execution}: We retrieved the original LintQ warnings for the 43 real-world files from the primary reports. For the 12 synthetic files, we executed the original rule-based \lintq to generate the baseline warnings.
    \item \textbf{\approachCoT Inference}: The 55 files were analyzed by \approachCoT to generate warnings. 
    \item \textbf{\approachRAG Inference}: The Evaluation Corpus was processed by the retrieval-enhanced approach to generate warnings.
\end{enumerate}
Finally, the outputs from \lintq, \approachCoT, and \approachRAG were compared against the established Ground Truth to calculate classification metrics.

\subsection{Evaluation Metrics}
\label{subsec:metrics}
To compute the metrics needed to study our RQs, we had to determine:
\begin{itemize}
    \item \textbf{True Positive (TP)}: The tool correctly generates a warning for a line of code containing the specific quantum programming problem.
    \item \textbf{False Positive (FP)}: The tool generates a warning for a line of code that does not contain that specific quantum programming problem.
    \item \textbf{False Negative (FN)}: The tool fails to generate a warning for a line that contains a quantum programming problem, as confirmed by the ground truth.
\end{itemize}

We used TP, FP, and FN to calculate the following metrics:
\begin{equation}
\text{Precision} = \frac{\mathit{TP}}{\mathit{TP} + \mathit{FP}}
\end{equation}
It estimates the correctness/reliability of the warnings issued by a given approach. A high precision value implies that developers will waste less time investigating false alarms.

\begin{equation}
\text{Recall} = \frac{\mathit{TP}}{\mathit{TP} + \mathit{FN}}
\end{equation}
It assesses the approach's completeness (i.e., sensitivity). A high recall value indicates the tool misses very few real bugs.

\begin{equation}
\text{F1-score} = 2 \cdot \frac{\text{Precision} \cdot \text{Recall}}{\text{Precision} + \text{Recall}}
\end{equation}
It provides a single, balanced assessment of the approach's overall effectiveness
by computing the harmonic mean of precision and recall. It accounts for the inherent trade-off between overwhelming the developer with false warnings (low precision) and missing critical quantum programming problems entirely (low recall). The higher its value, the better.

\section{Results} \label{sec:results}

This section presents the results of our empirical assessment and some complementary results to better understand the contribution of our research. Specifically, Table~\ref{tab:overall_results} summarizes the findings across the 55 files forming our Evaluation Corpus. The results suggest that both \approachCoT and \approachRAG substantially outperform the \lintq baseline in identifying quantum programming problems. In detail, \approachCoT achieved the highest recall value (0.96) and overall F1-score (0.70), demonstrating its capacity to identify almost every injected and real quantum programming problem in our corpus, keeping FNs to a minimum (3). Meanwhile, the integration of the semantic knowledge base in \approachRAG led to the highest precision value (0.56), minimizing the number of FPs reported compared to the base LLM approach, albeit at a cost in recall.

\begin{table}[]
    \centering
    \caption{Overall Performance Comparison across all Problem Types.}
    \label{tab:overall_results}
    \begin{tabular}{lcccccc}
        \toprule
        \textbf{Approach} & \textbf{TP} & \textbf{FP} & \textbf{FN} & \textbf{Precision} & \textbf{Recall} & \textbf{F1-Score} \\
        \midrule
        \lintq (Baseline) & 30 & 47 & 40 & 0.39 & 0.43 & 0.41 \\
        \approachCoT & 67 & 54 & 3 & 0.55 & \textbf{0.96} & \textbf{0.70} \\
        \approachRAG & 60 & 47 & 10 & \textbf{0.56} & 0.86 & 0.68 \\
        \bottomrule
    \end{tabular}
\end{table}

\subsection{RQ1: How do \approachCoT and \approachRAG compare to \lintq in terms of effectiveness, correctness, and completeness in detecting quantum programming problems?}

As far as the {effectiveness} is concerned, both LLM-based approaches outperform the baseline \lintq (see Table~\ref{tab:overall_results}). \approachCoT achieves the highest F1-score of 0.70, and \approachRAG closely follows with 0.68, whereas the rule-based \lintq baseline only reaches 0.41. This suggests that transitioning from static, manually crafted queries to the flexible reasoning capabilities of LLMs provides a massive uplift in overall detection capability.

In terms of correctness, the integration of the semantic knowledge base in \approachRAG proves beneficial, achieving the highest overall precision value of 0.56. This means the RAG-enhanced approach minimizes the number of FPs reported. \approachCoT achieves a similar precision value of 0.55, but both represent a substantial improvement over the 0.39 precision value of \lintq. We attribute this increased correctness to the CoT prompting and, for the RAG variant, to the one-shot learning examples that help ground the LLM's reasoning and reduce hallucinations.

Finally, in terms of {completeness}, \approachCoT demonstrates an outstanding capability of retrieving almost every quantum programming problem in the corpus, reaching a peak recall value of 0.96 with only 3 FNs. In comparison, \approachRAG achieves a recall value of 0.86, and the baseline falls short at 0.43. While RAG slightly penalizes recall by introducing stricter semantic constraints from the retrieved snippets, both \approachCoT and \approachRAG are evidently far superior to the baseline in comprehensively identifying quantum programming problems.

\begin{tcolorbox}[colback=gray!10!white,colframe=gray!50!black,title=\textbf{Summary of RQ1}]
Both \approachCoT and \approachRAG substantially outperform the rule-based \lintq~baseline across all evaluated metrics. \approachCoT achieves the highest effectiveness (F1-score of 0.70) and completeness (Recall of 0.96), while \approachRAG provides the highest correctness (Precision of 0.56) by minimizing FPs.
\end{tcolorbox}
\subsection{RQ2: To what extent does \approachRAG enhance \approachCoT's ability to detect quantum programming problems with respect to correctness, completeness, and effectiveness?}

To answer RQ2, we compare the performance of \approachCoT with its RAG augmented counterpart, \approachRAG, both shown in Table~\ref{tab:overall_results}. As for the {correctness} of the detected quantum program problems, the introduction of the RAG architecture yielded an improvement, even if it does not seem significant. Indeed, the precision value increased from 0.55 to 0.56, actively reducing the aggregate number of FPs from 54 down to 47. By grounding the reasoning on manually verified data, it seems that the LLM becomes more calibrated. 
However, this gain in correctness incurs a penalty in {completeness}. The recall metric dropped from a near-perfect 0.96 in the \approachCoT configuration down to 0.86 for \approachRAG, with FNs increasing from 3 to 10. A plausible explanation is that the explicit nature of the retrieved one-shot snippets acts as an overly rigid template. 
In other words, the strict semantic constraints introduced by RAG blind the model to valid, yet structurally diverse, implementations of the same quantum programming problem.

In terms of effectiveness, the RAG integration marginally underperformed \approachCoT. The decrease in Recall outweighed the slight improvement in Precision, resulting in a minor decline in the overall F1-score from 0.70 to 0.68.

\begin{tcolorbox}[colback=gray!10!white,colframe=gray!50!black,title=\textbf{Summary of RQ2}]
The use of RAG introduces a trade-off between correctness and completeness in the detection of quantum programming problems. Its use slightly improves correctness (Precision increases from 0.55 to 0.56) at the cost of a reduced completeness of the detected program problems (Recall drops from 0.96 to 0.86). As a result, the overall effectiveness (0.68 vs. 0.70) slightly decreases, indicating that the gain in precision does not compensate for the loss in recall. Therefore, RAG enhances reliability but limits generalization, making it beneficial in scenarios where precision is prioritized, but less suitable when comprehensive detection is strongly required.

\end{tcolorbox}

\subsection{Additional Analyses}

\subsubsection{On the Impact of Multi-prompt and CoT }

For completeness, we performed a comparative analysis between \approachCoT and \approach introduced in prior work~\cite{shin2025}. \approach achieved a high completeness with a recall value of 0.99 (TP = 69, FN = 1). However, this result comes at the expense of correctness, as the precision value is only 0.32 due to an exceptionally high number of FPs (145). This outcome suggests that while single-prompt strategies are effective at identifying potential problems, they lack the necessary capability to distinguish between correct quantum programming code and quantum programming problems, resulting in an F1-score of 0.49.

The transition to a multi-prompt CoT pipeline (\approachCoT) marks a performance milestone. By guiding the LLM, we achieved an increase in Precision to 0.55, effectively reducing the number of FPs by 62.76\% (from 145 to 54). Remarkably, this refinement in correctness does not significantly compromise completeness, as the Recall remains high at 0.96. Consequently, the overall effectiveness (F1-score) rises to 0.70, representing a 30\% improvement over \approach. Based on the achieved results, similar results and considerations can also be done when comparing \approach and \approachRAG.

\subsubsection{Obfuscation and Data Leakage}
A significant validity threat in LLM-based software engineering research is data leakage---the possibility that the underlying model may have encountered our test corpus during its pre-training phase. To empirically address this, we conducted a preliminary experiment using a custom obfuscator, structurally mapping user-defined identifiers to random strings while preserving syntactic integrity and Qiskit API interactions. Testing a sample of 14 files under both obfuscated and clear conditions (as detailed in Table~\ref{tab:obfuscation_results}) revealed that \approachCoT maintained comparable problem detection capability regardless of identifier obfuscation. This suggests the model anchors its reasoning on the topological properties of the quantum circuit construction rather than memorized sequences due to data leakages. Conversely, the semantic representations within the RAG solution proved highly sensitive to obfuscation, disjointing the retrieval correlations (only 97 out of 140 retrieval queries successfully aligned with their un-obfuscated counterparts). Consequently, since obfuscation hampered retrieval metrics without yielding significant data leakage issues in \approachCoT execution, the primary experiment was conducted using un-obfuscated Qiskit programs to maximize the structural efficacy.

\begin{table}[]
    \centering
    \caption{Performance comparison between Obfuscated (OBF) and Non-Obfuscated (Clear) inputs on a sample of 14 files.}
    \label{tab:obfuscation_results}
    \begin{tabular}{lccc}
        \toprule
        \textbf{Configuration} & \textbf{TP} & \textbf{FP} & \textbf{FN} \\
        \midrule
        \approachRAG (OBF) & 4 & 0 & 8 \\
        \approachCoT (OBF) & 8 & 3 & 4 \\
        \midrule
        \approachRAG (Clear) & 5 & 0 & 7 \\
        \approachCoT (Clear) & 7 & 1 & 5 \\
        \bottomrule
    \end{tabular}
\end{table}

\subsubsection{RAG and Types of Quantum Programming Problems}

\begin{table*}[t]
\centering
\caption{Comparison between \approachCoT and \approachRAG by type of quantum programming problem.}
\label{tab:cot-vs-rag-bugtype}
\small
\setlength{\tabcolsep}{4.5pt}
\renewcommand{\arraystretch}{1.15}
\resizebox{0.85\textwidth}{!}{\begin{tabular}{lc|ccc|cccccc}
\toprule
\multirow{2}{*}{Quantum Problem} & \multirow{2}{*}{\begin{tabular}[c]{@{}c@{}}Number of examples  available \\ in RAG Knowledge base\end{tabular}} & \multicolumn{3}{c|}{\approachCoT}                               & \multicolumn{3}{c}{\approachRAG}                                  & \multicolumn{3}{c}{RAG Impact (RAG--COT)}                                                             \\ \cline{3-11} 
                                 &                                                                                                               & \multicolumn{1}{l}{Prec.} & \multicolumn{1}{l}{Rec.} & \multicolumn{1}{l|}{F1} & \multicolumn{1}{l}{Prec.} & \multicolumn{1}{l}{Rec.} & \multicolumn{1}{l|}{F1}   & \multicolumn{1}{l}{$\Delta$Prec.} & \multicolumn{1}{l}{$\Delta$Rec.} & \multicolumn{1}{l}{$\Delta$F1} \\ \hline
OpAfterTransp                    & 4                                                                                                             & 1.00                      & 1.00                     & 1.00                    & 1.00                      & 1.00                     & \multicolumn{1}{c|}{1.00} & 0.00                              & 0.00                             & 0.00                           \\
GhostCompose                     & 6                                                                                                             & 1.00                      & 1.00                     & 1.00                    & 1.00                      & 1.00                     & \multicolumn{1}{c|}{1.00} & 0.00                              & 0.00                             & 0.00                           \\
OldIdenGate                      & 10                                                                                                            & 1.00                      & 1.00                     & 1.00                    & 1.00                      & 1.00                     & \multicolumn{1}{c|}{1.00} & 0.00                              & 0.00                             & 0.00                           \\
CondWoMeas                       & 14                                                                                                            & 0.83                      & 1.00                     & 0.91                    & 0.64                      & 0.80                     & \multicolumn{1}{c|}{0.73} & -0.17                             & -0.20                            & -0.18                          \\
DoubleMeas                       & 15                                                                                                            & 0.64                      & 1.00                     & 0.80                    & 0.86                      & 1.00                     & \multicolumn{1}{c|}{0.92} & 0.19                              & 0.00                             & 0.12                           \\
MeasAllAbuse                     & 16                                                                                                            & 1.00                      & 1.00                     & 1.00                    & 1.00                      & 1.00                     & \multicolumn{1}{c|}{1.00} & 0.00                              & 0.00                             & 0.00                           \\
InsuffClasReg                    & 18                                                                                                            & 0.09                      & 0.64                     & 0.16                    & 0.22                      & 0.64                     & \multicolumn{1}{c|}{0.33} & 0.13                              & 0.00                             & 0.17                           \\
OversizedCircuit                 & 20                                                                                                            & 0.11                      & 0.33                     & 0.17                    & 0.11                      & 0.67                     & \multicolumn{1}{c|}{0.19} & 0.00                              & 0.33                             & 0.02                           \\
ConstClasBit                     & 25                                                                                                            & 0.42                      & 1.00                     & 0.59                    & 0.38                      & 1.00                     & \multicolumn{1}{c|}{0.55} & -0.04                             & 0.00                             & -0.04                          \\
OpAfterMeas                      & 38                                                                                                            & 0.69                      & 1.00                     & 0.82                    & 0.69                      & 0.72                     & \multicolumn{1}{c|}{0.71} & 0.0                               & -0.28                            & -0.11                          \\ \hline
\end{tabular}
}
\end{table*}

Table~\ref{tab:cot-vs-rag-bugtype} details the values for Precision, Recall, and  F1-score for both \approachCoT and \approachRAG across the ten different quantum programming problem categories. The differences in the achieved results between these approaches are also reported.  
The effect of introducing RAG varies significantly depending on the specific problem type. 
For certain categories, RAG provides a notable performance uplift: \textit{DoubleMeas} elevated its F1-score from 0.80 to 0.92 (due to improved precision), and \textit{InsuffClasReg} improved from 0.16 to 0.33 (also due to improved precision). 
Conversely, other problem categories showed a regression in F1-scores when implementing RAG. Specifically, in \textit{OpAfterMeas} F1-score drop from 0.82 to 0.71 (driven by a decrease in recall), and \textit{ConstClasBit} decreased from 0.59 to 0.55 (driven by a slight decrease in precision). 
Furthermore, several problem types, such as \textit{OpAfterTransp}, \textit{GhostCompose}, \textit{OldIdenGate}, and \textit{MeasAllAbuse}, maintained a perfect classification score (F1-score = 1.0) across both configurations.

These findings suggest that the effectiveness of RAG is highly dependent on the quantum programming problem. We can also postulate on the dimension of the knowledge base available for each problem: the sheer number of examples does not appear to be directly associated with better performance. In fact, some of the most populated categories (e.g., \textit{OpAfterMeas} with 38 examples) suffered deteriorations, while categories with very few examples (e.g., \textit{OpAfterTransp} with 4 examples) maintained perfect scores. This suggests that the quality and structural representation of the examples are more critical than their quantity.

\section{Discussion and Practical Implications} \label{sec:discussion}

Our study highlights a shift in the potential mechanisms for quality assurance within quantum software engineering. The substantial outperformance of LLM-based approaches over the state-of-the-art rule-based tool (\lintq) suggests that generative models can effectively function as ``expert linters''. Unlike static analyzers that rely on rigid, declarative queries, LLMs can dynamically adapt to the rapidly evolving scene of quantum computing programming, effectively transferring the maintenance burden from manual rule creation to prompt engineering and knowledge base curation. For practitioners and tool developers/vendors, this implies that integrating LLMs into IDEs or continuous integration pipelines could offer a more robust and adaptable first line of defense against quantum programming problems.

While RAG successfully grounded the LLM and improved the correctness of the detection of quantum program problems, it simultaneously constrained the model, causing it to dismiss valid quantum programming problems that deviated structurally from the retrieved templates. To some extent, this finding reveals a trade-off that could be of interest tp practitioners. That is, in a context where reducing FNs is the priority, the use of CoT and RAG (i.e., \approachRAG) could be preferable. Conversely, if high completeness is required to ensure no quantum programming problem slips through, the \approachCoT approach could be preferred. The results also suggest that when it comes to RAG for quantum programming type, bigger is not necessarily better. Success seems driven by the structural and syntactic clarity of the problems rather than the sheer volume of the knowledge base. This outcome is clearly relevant for researchers interested in understanding when well-structured examples are more valuable than several generic data points (thus favoring data curation over data collection).

Furthermore, our investigation into code obfuscation revealed that the base generative model accurately identifies quantum programming problems based on their topological and structural properties. This resilience is promising in a context where proprietary quantum algorithms might need to be obfuscated before being analyzed by external cloud-based LLMs. Unfortunately, the evidence gathered so far shows that the semantic embeddings utilized by the FAISS index are highly sensitive to obfuscation, severely degrading the retrieval phase. These considerations point toward the necessity of developing structural or AST-aware embedding techniques that can withstand syntactic obfuscation while preserving the semantic context of quantum operations. 

Overall, these considerations are not only clearly relevant to researchers but maybe of interest also to
tool developers/vendors developing resilient, privacy-preserving automated analysis tools and advancing structural code representation models for quantum software development.

\section{Threats to Validity}
In the following, we present the possible threats that could affect the validity of the obtained results.
\paragraph{Internal Validity}
The performance of \approachCoT and \approachRAG heavily depends on the reasoning capabilities of the used LLM and its non-determinism. In our study, we utilized a single state-of-the-art GPT-5 model for all inference tasks. We acknowledge that different LLMs might exhibit varying degrees of proficiency, potentially altering the observed effectiveness. As for the non-determinism, we used the default configuration for the LLM.

Finally, to construct the Evaluation Corpus and ensure adequate representation of extremely rare problems, we injected synthetic quantum programming problems into a subset of files. While these injections structurally mirrored documented quantum programming problems, they might not perfectly replicate the complex topologies of organic, developer-induced problems.
Finally, the semantic retrieval mechanism in \approachRAG relies exclusively on the \texttt{text-embedding-3-large} model. The choice of the embedding model and distance metric directly governs the quality of the injected one-shot examples. A different embedding strategy could potentially impact the results.

\paragraph{External Validity}
While the abstract generative nature of LLMs theoretically allows for the analysis of other quantum languages such as Cirq or Q\#, we cannot claim that our quantitative results generalize beyond the Qiskit ecosystem~\cite{Qiskit}.

Finally, we cannot guarantee that our results perfectly generalize to all quantum programs developed in the wild. Although our stratified Evaluation Corpus consists of 55 files covering all ten recognized quantum programming problem families, it may not encapsulate the entirety of syntactic and structural variations present in larger-scale quantum repositories. To partially mitigate this threat, we provide a replication package available online~\cite{replication}.

\paragraph{Construct Validity}
A significant threat to construct validity lies in how the ``ground truth'' was established. Specifically, our manual validation was performed by a single human annotator. This introduces inherent risks associated with the subjective interpretation of quantum programming problems. To mitigate this threat, the annotator (the first author) strictly adhered to the formal problem definitions and code patterns exhaustively documented in prior literature~\cite{PaltenghiP24}. In case of doubts, which never occurred, we planned for the annotator to consult with one of the other authors to resolve them.

While Precision, Recall, and F-score are standard metrics, they treat all quantum programming problems equally. That is, there could be quantum programming problems less critical (e.g., OldIdenGate) than others because they are less detrimental, and we did not take this aspect into account.

\section{Conclusion} \label{sec:Conclusion}

We explored the feasibility and potential of LLMs to act as expert linters for quantum program analysis. Specifically, we introduced and assessed \approachCoT and \approachRAG, two variants of the \approach proposed in prior work~\cite{shin2025}, by integrating a multi-prompt Chain-of-Thought (CoT) pipeline with and without a Retrieval-Augmented Generation (RAG) extension grounded in manually verified quantum programming problem examples. 
To evaluate \approachCoT and \approachRAG, we applied them to a comprehensive, stratified Evaluation Corpus of 55 quantum programs. Our empirical results clearly indicate that transitioning from rigid, deterministic queries to flexible LLM-based reasoning yields a massive uplift in detection capabilities. While state-of-the-art static analysis tools like \lintq inherently struggle with adaptability, the foundational \approachCoT achieved the highest overall capability (F1-score of 0.70) by effectively generalizing across structural nuances. Concurrently, \approachRAG demonstrated the highest correctness metric (Precision of 0.56) by seamlessly utilizing context-informed, one-shot examples to reject misleading patterns and actively minimize FP alerts.

Looking ahead, our study sets the stage for the following future research directions:
\begin{itemize}\item \textbf{AST-Aware Embeddings:} Our preliminary testing showed that standard semantic embeddings (like those in FAISS) are highly sensitive to syntactic obfuscation. Future efforts must focus on developing structural or Abstract Syntax Tree (AST)-aware embedding techniques that withstand code obfuscation while preserving the deep semantic context of quantum operations.
\item \textbf{Human-Centered Interactions:} As LLM-powered linters mature into IDEs, evaluating the human dimension becomes paramount. Future human-based empirical studies should explore how developers perceive, interact with, and act on LLM-generated explanations compared with traditional linter outputs, and possibly explore hybrid systems that merge strict static tracking with generative recommendations.
\end{itemize}

\bibliographystyle{IEEEtran}

\end{document}